# Extent of stacking disorder in diamond

*Christoph G. Salzmann,*[a,]* *Benjamin J. Murray,*[b] *Jacob J. Shephard*[a]

[a] Department of Chemistry, University College London, 20 Gordon Street, London WC1H 0AJ, UK

[b] Institute for Climate and Atmospheric Science, School of Earth and Environment, University of Leeds, Leeds LS2 9JT, UK

* Corresponding author. E-mail: c.salzmann@ucl.ac.uk (Christoph G. Salzmann)

**Abstract**

Hexagonal diamond has been predicted computationally to display extraordinary physical properties including a hardness that exceeds cubic diamond. However, a recent electron microscopy study has shown that so-called hexagonal diamond samples are in fact not discrete materials but faulted and twinned cubic diamond. We now provide a quantitative analysis of cubic and hexagonal stacking in diamond samples by analysing X-ray diffraction data with the DIFFaX software package. The highest fractions of hexagonal stacking we find in materials which were previously referred to as hexagonal diamond are below 60%. The remainder of the stacking sequences are cubic. We show that the cubic and hexagonal sequences are interlaced in a complex way and that naturally occurring Lonsdaleite is not a simple phase mixture of cubic and hexagonal diamond. Instead, it is structurally best described as stacking disordered diamond. The future experimental challenge will be to prepare diamond samples beyond 60% hexagonality and towards the so far elusive 'perfect' hexagonal diamond.

## 1. Introduction

The common form of diamond is cubic. Yet, a metastable hexagonal polymorph has been identified in fragments of the Canyon Diablo meteorite from the northern Arizona desert and other impactites.[1, 2] To honour the achievements of the crystallographer Kathleen Lonsdale hexagonal diamond has been named Lonsdaleite.[1] Synthetically, hexagonal diamond can be prepared, for example, by heating graphite in the 15 – 20 GPa range.[3-8] There is considerable current interest in the targeted preparation and physical properties of hexagonal diamond as it has been predicted to display superior mechanical properties, such as hardness and compressive strength, compared to its cubic counterpart.[9, 10] Furthermore, cubic and hexagonal diamonds are expected to have different band gaps and dielectric properties.[11]

Cubic and hexagonal diamond both consist of $sp^3$ hybridized and therefore tetrahedrally-bonded carbon atoms. Both allotropes contain puckered layers of carbon atoms with six-membered rings in the armchair configuration. The difference between cubic and hexagonal diamond lies in how these layers are stacked on top of each other to build up the three-dimensional crystal structure (*cf.* Fig. 1). In cubic diamond, identical layers are stacked on top of each other with a shift half way across the diagonal of a six-membered ring. In hexagonal diamond on the other hand each layer is the mirror image of the previous layer.[12] The structural consequence of these different stacking recipes is that the six-membered rings *linking* the various layers are in the armchair conformation in cubic diamond but boat-type in hexagonal diamond. Consequently, cubic diamond consists of only armchair rings whereas hexagonal diamond is a 50:50 mixture of rings in the armchair and boat conformations, respectively.

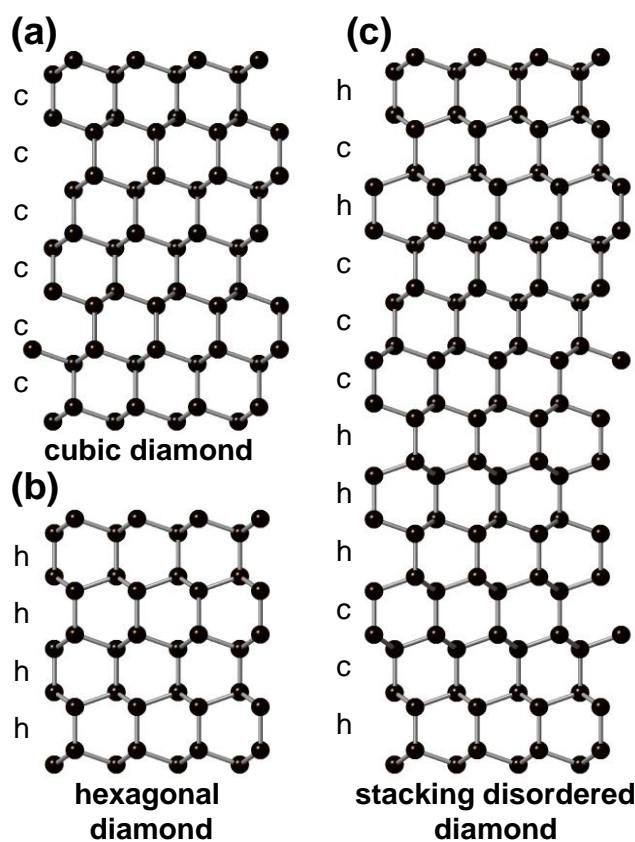

**Figure 1.** Crystal structure projections along the hexagonal *a* axis for (a) cubic diamond, (b) hexagonal diamond and (c) stacking disordered diamond.

The diamond structures are isostructural with ice if only the oxygen atoms in ice are considered. For ice, the hexagonal polymorph (ice I*h*) is the most stable at ambient pressure and a metastable cubic form (ice I*c*) has been thought to exist.[13] The field has progressed in recent years and it has been shown that what was previously considered to be cubic ice is in fact stacking disordered ice (ice I*sd*) containing variable fractions of both hexagonal as well as cubic stacking.[14-18] The 'perfect' cubic ice, containing only cubic stacking, has so far not been identified. As predicted for diamond, the differences in stacking have pronounced effects on the physical properties of ice including the vapour pressure,[19] crystal shapes,[20] spectroscopic properties[21] and potentially surface chemistry.[22]

Originally, it was assumed that hexagonal and cubic diamonds often coexist which helped understanding the observed X-ray diffraction data.[1, 2] Yet, very recently it has been shown by using high-resolution electron microscopy that what has been originally classified as hexagonal diamond is in fact not a discrete material but "faulted and twinned cubic diamond".[8]

Following our recent work on ice we show here how stacking disorder can be characterised quantitatively in diamond samples on the basis of X-ray diffraction data using the DIFFaX software package.[23] We thereby take so-called memory effects, where the stacking depends on the previous stacking history, into account. The aim is to fully describe the stacking disorder in the various samples of hexagonal diamond that have been made so far and to show which experimental recipes lead to the most hexagonal stacking in diamond.

## 2. Methodology

Diffraction patterns of a cubic diamond sample in a 0.6 mm glass capillary were collected on a Stoe Stadi-P diffractometer (Cu K$\alpha$, $\lambda$=1.540598 Å) with a Mython area detector. Additional diffraction data was taken from refs [6-8]. All diffraction patterns were background-corrected using shifted Chebyshev polynomials. For the calculation of diffraction patterns of stacking disordered diamond structure we used the DIFFaX software package[23] as previously employed for ice samples.[16, 18] To refine the $a$ and $c$ lattice constants, stacking probabilities, profile parameters ($u$, $v$, $w$ and GL ratio) and zero-shift we used our own MCDIFFaX programme which embeds DIFFaX in a least-squares environment.[1] A typical refinements started with the optimisation of the lattice constants and the peak profile parameters. This was followed by refining the various stacking probabilities. Typically, several tens of thousands of individual DIFFaX calls were needed before the refinements converged. To prevent false minima, MCDIFFaX uses a Monte-Carlo-type parameter that allows a defined fraction of unfavourable moves to take place.

---

[1] Salzmann CG, www.ucl.ac.uk/chemistry/research/group_pages/salzmann_group

## 3. Results and Discussion

The characteristics of stacking disorder in diamond are conveniently summarised with a so-called stackogram as shown in Figure 2(a). In such a diagram, 1st order memory effects are taken into account; these are described with two independent stacking probabilities. The probability of cubic stacking following a cubic event is given by $\Phi_{cc}$ whereas $\Phi_{hc}$ describes the probability of a cubic event after hexagonal stacking. Since the stacking can only be either cubic or hexagonal it follows that $\Phi_{ch} = 1 - \Phi_{cc}$ and $\Phi_{hh} = 1 - \Phi_{hc}$. The stackogram is a plot of $\Phi_{cc}$ against $\Phi_{hc}$ and the corresponding dependent probabilities, $\Phi_{ch}$ and $\Phi_{hh}$, have also been included. The four corners of this diagram define the end-member states. If $\Phi_{cc}$ and $\Phi_{hc}$ are both zero the probability for cubic stacking is zero and consequently this corner represents hexagonal diamond. In turn, cubic diamond is defined by $\Phi_{cc}$ and $\Phi_{hc}$ both equal to one. A physical mixture of cubic and hexagonal diamond implies $\Phi_{cc} = 1$ and $\Phi_{hc} = 0$ (*i.e.* $\Phi_{hh} = 1$) whereas the strictly alternating (hc)$_x$ polytype is defined by $\Phi_{cc} = 0$ (*i.e.* $\Phi_{ch} = 1$) and $\Phi_{hc} = 1$. The latter two cases are extremes of 1st order memory effects. Either once in a certain stacking sequence it has to continue infinitively or it must be strictly alternating. Depending on the exact location on the stackogram the 1st order memory effects will be more or less pronounced. In fact, along the diagonal connecting the hexagonal diamond corner with cubic diamond $\Phi_{cc}$ and $\Phi_{hc}$ are equal which means no memory effects. The stacking is purely random and independent of the previous stacking history.

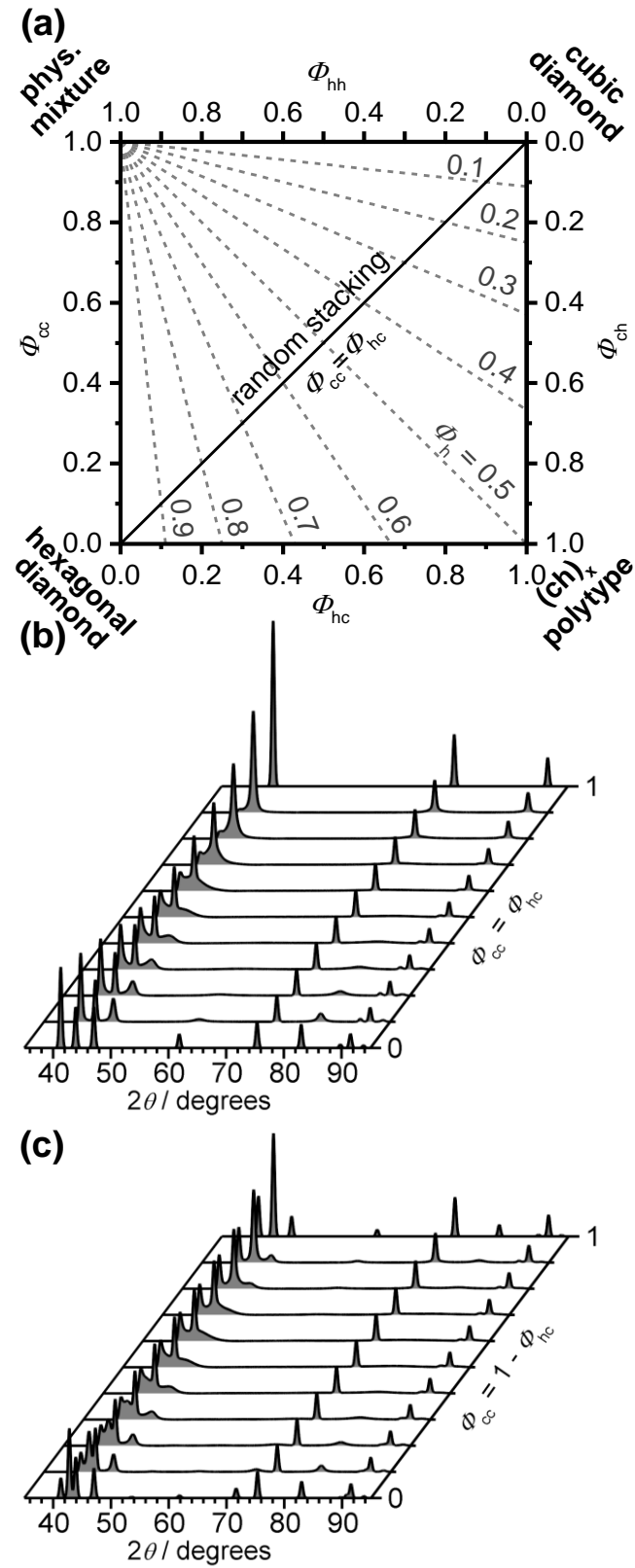

**Figure 2.** (a) Stackogram used for the structural description of stacking disorder in diamond including 1st order memory effects. The black solid line indicates structures with random stacking whereas the dashed lines describe structures with a constant hexagonality. (b,c)

Calculated powder diffraction patterns (Cu Kα) along the random stacking as well as the 0.5 hexagonality line.

A quantity of primary interest to describe a given stacking disordered material is the fraction of cubic stacking which is also called the 'cubicity'.[17] This parameter, $Φ_c$, can be calculated from the 1st order stacking probabilities according to

$$Φ_c = \frac{Φ_{hc}}{Φ_{hc} + Φ_{ch}} .  \quad [1]$$

For diamond in particular, the interest lies more in the fraction of hexagonal stacking and it is therefore sensible to define the 'hexagonality' of a sample, $Φ_h$, which is $1 − Φ_c$. The stackogram in Figure 2(a) shows lines of constant hexagonality which originate from the corner describing the physical mixture of polymorphs. Strictly speaking, the hexagonality is not defined in this corner since no switching between the two kinds of stacking is allowed. In fact, division by zero would take place in equation 1.

Figure 2(b) shows calculated X-ray diffraction patterns using DIFFaX along the random-stacking line ($Φ_{cc} = Φ_{hc}$) from hexagonal to cubic diamond. Hexagonal diamonds displays the characteristic 'trident' between 40 and 50 degrees also observed for ice I$h$ in a different angle range. The central (002) peak persists as the fraction of cubic stacking increases. Yet, broad and asymmetric diffraction features develop on both the high as well as low angle side which are the hallmarks of stacking disorder. Moving away from defined spots in reciprocal space for perfectly crystalline materials the stacking disorder leads to 'streaking'. The crystallographic rule is that only Bragg peaks (*hkl*) where $(h − k)/3$ is not an integer number are affected by stacking disorder.[24] Generally, Bragg peaks with greater values of *l* are more affected by stacking disorder.[25] For the cubic end member, one peak remains in the 40 − 50 degrees angle range which is the (111) reflection of the cubic system. The effects of 1st order memory effects are shown in Figure 2(c) where calculated diffraction patterns along the 0.5 hexagonality line are shown. Again, diffuse scattering is only observed for the stacking disordered states, *i.e.* neither for the physical mixture nor the strictly alternating (hc)$_x$ polytype.

We next analyse various experimental diamond diffraction patterns with MCDIFFaX with the aim to pinpoint their respective positions in the stackogram. For this, we additionally include 2$^{nd}$ order memory effects which are described by four independent stacking probabilities $\Phi_{ccc}$, $\Phi_{hcc}$, $\Phi_{chc}$ and $\Phi_{hhc}$. The corresponding 1$^{st}$ order probabilities and consequently the corresponding hexagonality are calculated according to

$$\Phi_{cc} = 1 - \frac{\Phi_{cch}}{\Phi_{cch} + \Phi_{hcc}} \text{ and} \tag{2}$$

$$\Phi_{hc} = \frac{\Phi_{hhc}}{\Phi_{hhc} + \Phi_{chh}}. \tag{3}$$

Figure 3(a) shows a diffraction pattern of a natural diamond sample with ~1 µm sized crystallites. The diffraction pattern is fitted very well with $\Phi_c = 1$ and no signs of stacking disorder have been detected. As previously for ice, we suggest that a diamond sample is considered to be stacking disordered if it contains more than one percent of the minor stacking component.[16, 18] Below this limit it is probably more appropriate to speak of stacking faults, although this boundary is arbitrary.

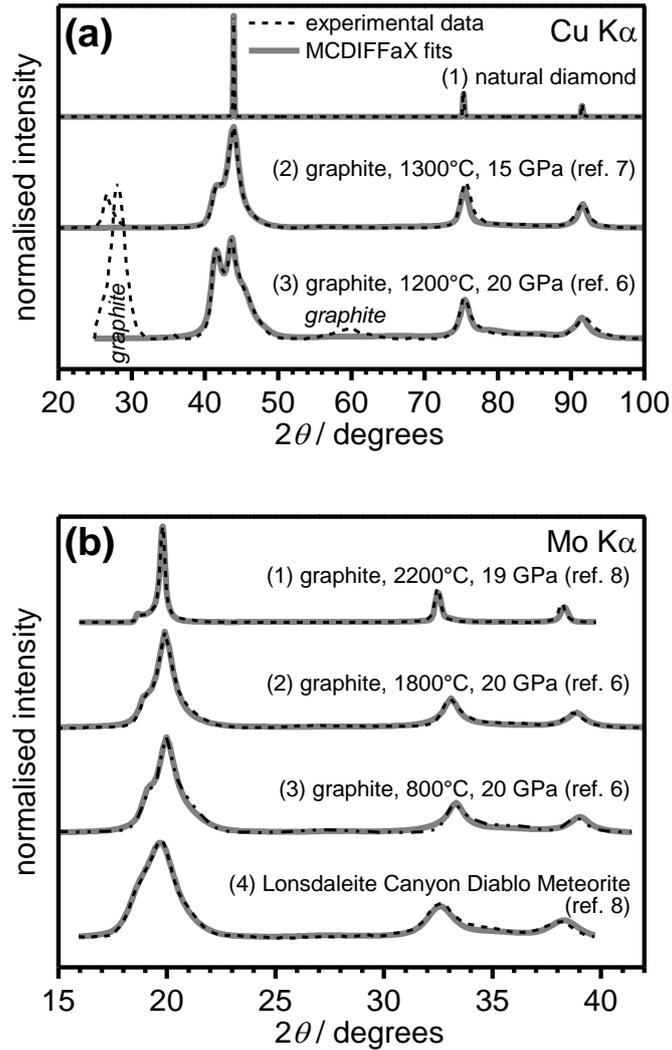

**Figure 3.** Experimental powder diffraction data of various diamond samples (dashed black lines) fitted with MCDIFFaX (thick grey lines). The X-ray sources are Cu Kα ($\lambda$=1.540598 Å) for (a) and Mo Kα ($\lambda$=0.7093165 Å) for (b).

Several powder diffraction patterns of stacking disordered diamond from the literature were analysed next.[6-8] For this kind of analysis it is essential to have as much diffraction data above the 'trident' angle range as possible and literature patterns where only this range was shown were not analysed. The stacking disordered diamond samples were obtained by heating graphite to the pressures and temperatures indicated in Fig. 3.[6-8] The issue of preferred orientation was addressed in ref. [6] and we were able to fit all their diffraction data. The data from Fig. 3 in ref. [7] on the other hand was more problematic in this respect and we have only been able to fit one of their diffraction patterns. Comparing the diffraction data from ref. [7] with the calculated patterns in Fig. 2 clearly shows that preferred orientation effects must have had an effect on some of the peak intensities. This is in fact not surprising considering that highly-oriented pyrolytic graphite was used as the starting material in ref. [7] whereas in ref. [6], for comparison, fine graphite powders were used.

Pattern (4) in Fig. 3(b) is that of a Lonsdaleite sample recovered from the Canyon Diablo meteorite after dissolving the iron parts in dilute hydrochloric acid.[8] The diffraction features of the Lonsdaleite sample are the broadest of all diffraction patterns shown in Fig. 3 which points towards smaller domain sizes.

The $1^{st}$ order stacking probabilities obtained from fitting the diffraction data of the various diamond samples are shown in Fig. 4. Several important points can now be made: (1) No sample displays a hexagonality greater than 0.6. This quantitative statement supports the conclusion from ref. [8] that pure hexagonal diamond has so far not been prepared. (2) All analysed diamond samples lie above the random stacking line in the stackogram. As discussed earlier, this indicates a propensity in diamond to stay in a given stacking sequence rather than to alternate between hexagonal and cubic stacking. In this respect, the situation is similar as previously observed for ice I.[16, 18] (3) All samples are quite far away from the physical mixture corner with the natural Lonsdaleite sample being the closest. This underpins the fact that Lonsdaleite should not be regarded as a mixture of cubic and hexagonal diamonds but stacking disordered diamond instead. (4) In ref. [8] it was stated that Lonsdaleite is "faulted and twinned cubic diamond". However, as it can be seen in Fig. 4, the Lonsdaleite sample is close to the 0.5 hexagonality line and therefore quite far away from the cubic diamond corner. This means that Lonsdaleite is in our opinion best described as stacking disordered diamond. (5) It is inaccurate to describe Lonsdaleite as hexagonal diamond.

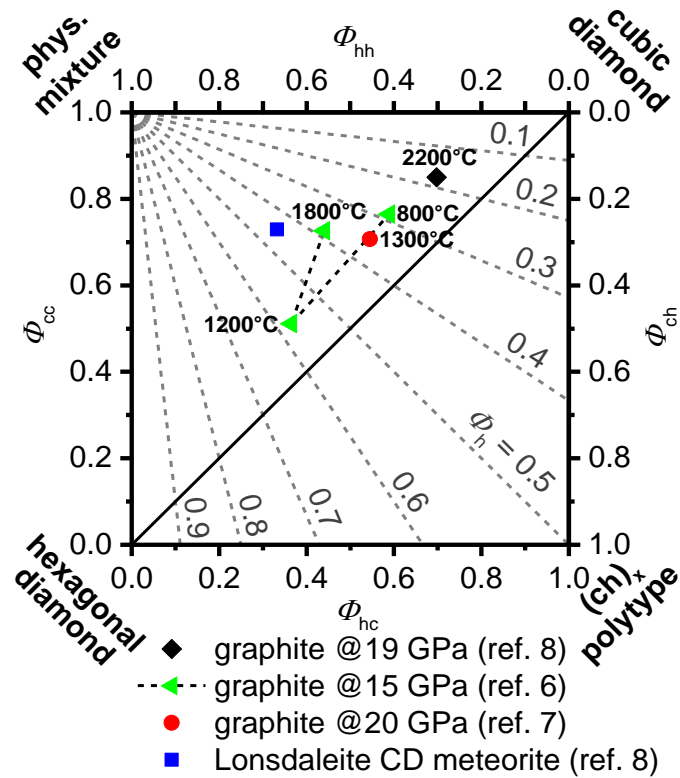

**Figure 4.** Stackogram with the 1$^{st}$ order memory stacking probabilities obtained from MCDIFFaX fits of the diffraction data shown in Figure 3.

In summary, we have shown that quantitative information about the stacking disorder in diamond samples can be obtained from X-ray diffraction patterns. It is interesting to note in this context that the highest hexagonality was obtained for a sample which was pressure-annealed at 1200°C, a temperature between two other experiments where more cubic samples were obtained.[6] This seems to highlight the very complex kinetics of the formation of diamond from graphite. The experimental challenge is now to prepare diamond samples well beyond the 0.6 hexagonality line and towards the 'perfect' hexagonal diamond which has been predicted computationally to be a highly extraordinary material.[9-11, 26]

**Acknowledgments**

We thank the Royal Society (CGS, UF100144) the Leverhulme Trust (RPG-2014-04) and the European Research Council (FP7, 240449 ICE) for financial support.